\documentclass[aps,prd,onecolumn,groupedaddress,showpacs,nofootinbib,amssymb]{revtex4}
\usepackage[dvips]{graphicx}
\usepackage{amssymb}
\usepackage{amsmath}
\usepackage{graphicx,,color}
\usepackage{amsfonts}
\usepackage{bm}
\usepackage{cancel}
\usepackage{comment}

\newcommand\be{\begin{equation}}
\newcommand\ee{\end{equation}}

\allowdisplaybreaks[4]

\begin{document}

\title{Power-Law $f(R)$ Gravity Corrected Canonical Scalar Field Inflation}
\author{V.K. Oikonomou,$^{1,2}$\,\thanks{v.k.oikonomou1979@gmail.com}}
\affiliation{$^{1)}$Department of Physics, Aristotle University of Thessaloniki, Thessaloniki 54124, Greece\\
$^{2)}$ Laboratory for Theoretical Cosmology, Tomsk State
University of Control Systems and Radioelectronics, 634050 Tomsk,
Russia (TUSUR)}

\tolerance=5000

\begin{abstract}
The effective inflationary Lagrangian is a prominent challenge for
theoretical cosmologists, since it may contain imprints of the
quantum epoch. In view of the fact that higher order curvature
terms might be present in the effective inflationary Lagrangian,
in this work we introduce the theoretical framework of power-law
$f(R)$ gravity corrected canonical scalar field inflation, aiming
to study the inflationary dynamics of this new framework. The main
characteristic of this new theoretical framework is the dominance
of a power-law $f(R)$ gravity term $\sim R^n$, with $1<n<2$,
compared to the Einstein-Hilbert term $\sim R$. In effect, the
field equations are controlled by the $R^n$ term in contrast to
the Einstein-Hilbert canonical scalar theory. We extract the
slow-roll field equations and we calculate the slow-roll indices
of the resulting theory which acquire quite elegant final form,
when the slow-roll conditions hold true. Accordingly, we examine
quantitatively the inflationary phenomenological implications of
the theoretical framework we introduced, by choosing simple
hybrid-like scalar field potentials. As we evince the resulting
theory is in good agreement with the latest Planck data for a wide
range of the free parameters of the model. Thus our theoretical
framework makes possible to obtain viable inflationary theories,
which otherwise would be non-viable, such as the simple power-law
$f(R)$ gravity model or the simple power-law scalar model.
\end{abstract}


\maketitle

\section{Introduction}

After the primordial quantum epoch, the Universe is believed to
have inflated in an abrupt way, during an era which is known as
the inflationary era. At the beginning of the inflationary era,
the Universe was hot, described by a four dimensional spacetime
metric and was more or less in a classical state, with possibly
only remnants of the full high energy quantum theory making their
presence in the effective inflationary Lagrangian. To date, the
most well-known description of inflation is that of a scalar
field, the inflaton, which controls the inflationary and early
post-inflationary era
\cite{inflation1,inflation2,inflation3,inflation4}. The
inflationary scenario solves quite prominent problems of the
standard Big Bang scenario, such as the flatness problem and the
horizon problem, and several models of scalar field inflation are
robust towards the constraints imposed by the latest Planck data
\cite{Akrami:2018odb}. Nevertheless, it is not certain if
inflation even occurred, since a direct evidence of the
inflationary era would be provided by the observation of $B$-modes
(curl) in the Cosmic-Microwave-Background radiation
\cite{Kamionkowski:2015yta}. There are appealing alternative
scenarios that may also describe the primordial epoch in a
successful way, such as the bouncing cosmology scenarios
\cite{Brandenberger:2012zb,Brandenberger:2016vhg,Battefeld:2014uga,Novello:2008ra,Cai:2014bea,deHaro:2015wda,Lehners:2011kr,Lehners:2008vx,Cheung:2016wik,Cai:2016hea,Odintsov:2020vjb},
and in some cases a dark energy era might be combined with a
primordial bounce \cite{Odintsov:2020vjb}, however all these
scenarios, including the inflationary scenario, are for the moment
candidate proposals for the primordial epoch of our Universe, and
only constraints can be imposed on them.

Scalar field inflation strongly relies on the existence of a
primordial scalar field, the inflation, which effectively controls
the dynamical evolution of the Universe. However, to date, only
the only scalar field that has ever been observed is the Higgs
scalar \cite{Aad:2012tfa}, thus it is compelling to explain how
the inflation plays some fundamental role in the high energy
physics, and answer the question whether the inflaton is directly
related to the Higgs particle or even if the Higgs particle is the
inflaton itself. Apart from this, a vital question that should be
answered in a concrete way is the effective Lagrangian of
inflation itself. Since at the beginning of inflation, the
Universe emerged classical from a quantum epoch, it is possible
that the quantum epoch left its imprints in the classical
effective inflationary Lagrangian. Several scenarios that predict
corrections to the canonical scalar field Lagrangian exist in the
literature, such as the Einstein-Gauss-Bonnet corrections
\cite{Hwang:2005hb,Nojiri:2006je,Cognola:2006sp,Nojiri:2005vv,Nojiri:2005jg,Satoh:2007gn,Bamba:2014zoa,Yi:2018gse,Guo:2009uk,Guo:2010jr,Jiang:2013gza,Kanti:2015pda,vandeBruck:2017voa,Kanti:1998jd,Pozdeeva:2020apf,Fomin:2020hfh,DeLaurentis:2015fea,Chervon:2019sey,Nozari:2017rta,Odintsov:2018zhw,Kawai:1998ab,Yi:2018dhl,vandeBruck:2016xvt,Kleihaus:2019rbg,Bakopoulos:2019tvc,Maeda:2011zn,Bakopoulos:2020dfg,Ai:2020peo,Odintsov:2019clh,Oikonomou:2020oil,Odintsov:2020xji,Oikonomou:2020sij,Odintsov:2020zkl,Odintsov:2020sqy,Odintsov:2020mkz,Easther:1996yd,Antoniadis:1993jc,Antoniadis:1990uu,Kanti:1995vq,Kanti:1997br},
which are also string theory originating. Another possibility is
that at high curvatures, higher order curvature terms may
eventually be present in the effective inflationary Lagrangian and
these terms can play a dominant role in the inflationary dynamics
of simple scalar field inflation. In this line of research, in
this paper we shall assume that power-law $f(R)$ gravity terms are
present in the effective inflationary Lagrangian of a canonical
scalar field, and we shall examine the dynamical evolution of the
scalar field in the presence of these power-law $f(R)$ gravity
terms of the form $\sim R^{n}$. The $R^n$ term will dominate
during inflation over the Einstein-Hilbert term $\sim R$, and
consequently, the dynamical evolution of the canonical scalar
field will be affected in a crucial way. We shall assume that the
slow-roll conditions holds true and eventually the whole canonical
scalar field framework will be changed due to the presence of the
$\sim R^n$ terms in the inflationary Lagrangian of the scalar
field. We shall extract the slow-roll field equations and the
corresponding slow-roll indices, and we present a set of quite
elegant analytic expressions for the slow-roll indices and the
corresponding observational indices of inflation, namely the
spectral index of the primordial scalar curvature perturbations
and of the tensor-to-scalar ratio. The reader will recognize a
combination of well known expressions for the observational
indices corresponding to pure scalar field inflation and pure
$f(R)$ gravity inflation. Of course, $f(R)$ gravity by itself can
describe in a successful way the inflationary era
\cite{reviews1,reviews2,reviews3,reviews4,dimopoulos}, hence our
approach will provide a unified effect of scalar field theory and
of $f(R)$ gravity. The motivation for this is simple, by thinking
the late-time evolution of our Universe, which cannot be described
by Einstein-Hilbert gravity, even when scalar fields are present.
This is due to the fact that the effective equation of state
parameter is marginally allowed observationally to cross the
phantom divide line, as is indicated by the latest Planck data on
the cosmological parameters \cite{Aghanim:2018eyx}. Such a
behavior cannot be harbored by general relativity, since the
cosmological constant yields an equation of state parameter which
is exactly equal to minus one. In addition, even if someone uses
scalar fields to describe the late-time era, an equation of state
parameter which crosses the phantom divide line requires a phantom
scalar field \cite{Caldwell:2003vq}, which is basically
problematic by itself, being an instability. Thus, it is highly
possible that a single scalar field may not be enough to describe
the full effective inflationary Lagrangian. To our opinion,
Gauss-Bonnet terms and $f(R)$ gravity terms are the most prominent
candidates that may accompany the scalar field in the full
effective inflationary Lagrangian.


\section{Canonical Scalar Field Inflation in the Presence of $f(R)$ Gravity}

The canonical scalar field action in the presence of $f(R)$
gravity terms has the following form,
\begin{equation}\label{action1dse}
\mathcal{S}=\int
\mathrm{d}^4x\sqrt{-g}\left(\frac{R+\lambda\left(\frac{R}{R_0}\right)^n}{2\kappa^2}-\frac{1}{2}g^{\mu
\nu}\partial_{\mu}\varphi\partial_{\nu}\varphi-V(\varphi)\right),
\end{equation}
where $\kappa^2$ stands for $\kappa^2=8\pi G=\frac{1}{M_p^2}$ and
 $M_p$ is the reduced Planck mass. Also $\lambda$ and $R_0$ are
 dimensionful parameters both having dimensions $[m]^2$ in natural
 units, and we shall assume in the end that
 $\lambda=\frac{\tilde{\lambda}}{\kappa^2}$ and
 $R_0=\frac{1}{\kappa^2}$, where $\tilde{\lambda}$ is a
 dimensionless constant. In addition, initially we shall assume
 that the parameter $n$ will be in the range $1<n<2$, but as we
 show shortly, $n$ must be chosen in the range $3/2<n<2$ for
 the consistency of the slow-roll approximation. Assuming that the spacetime metric is
 that of a flat Friedmann-Robertson-Walker (FRW) metric of
the form,
\begin{equation}
\label{JGRG14} ds^2 = - dt^2 + a(t)^2 \sum_{i=1,2,3}
\left(dx^i\right)^2\, ,
\end{equation}
with $a(t)$ being the scale factor, upon varying the gravitational
action (\ref{action1dse}) with respect to the metric tensor, we
obtain the field equations, which read,
\begin{equation}
\label{motion1a} \centering 3H^2f_R=\frac{Rf_R-f}{2}-3H\dot{f}_R+
\kappa^2\left( \frac{1}{2}\dot\varphi^2+V(\varphi)\right)\, ,
\end{equation}
\begin{equation}
\label{motion2a} \centering -2\dot H f_R=
\kappa^2\dot\varphi^2+\ddot{f}_R-H\dot{f}_R\, ,
\end{equation}
\begin{equation}
\label{motion3a} \centering \ddot{\varphi}+3H\dot{\varphi}+V'=0\,
,
\end{equation}
where the ``prime'' denotes differentiation with respect to the
scalar field $\varphi$, and we shall keep this convention
hereafter, and $f_{R}=\frac{\mathrm{d}f}{\mathrm{d}R}$. During the
inflationary era, the slow-roll condition holds true for the
Hubble rate holds true, namely,
\begin{equation}\label{slowrollhubble}
\dot{H}\ll H^2,\,\,\,\ddot{H}\ll H\dot{H}\, ,
\end{equation}
thus the Ricci scalar and its derivative are approximately equal
to,
\begin{equation}\label{ricciscalarapprox}
R\sim 12H^2,\,\,\,\dot{R}\sim 24 H\dot{H}\, .
\end{equation}
In addition, even for the small scale inflation, the Hubble rate
is of the order $H_I\sim 10^{13}$GeV, thus in the Friedmann
equation (\ref{motion1a}), the power-law term $R^n$ dominates over
the Einstein-Hilbert term $R$, hence the Friedmann equation
(\ref{motion1a}) becomes approximately,
\begin{equation}
\label{motion1aapprox} \centering \frac{3\lambda n
H^2}{R_0}\left(\frac{R}{R_0}\right)^{n-1}=\frac{\lambda
(n-1)}{2}\left(\frac{R}{R_0}\right)^{n}-\frac{3H\lambda
n(n-1)}{R_0^n}R^{n-2}\left(24 H\dot{H}+6\ddot{H} \right)+
\kappa^2\left( \frac{1}{2}\dot\varphi^2+V(\varphi)\right)\, .
\end{equation}
Since $n>1$, and in view of the slow-roll conditions for the
Hubble rate (\ref{slowrollhubble}), the term $\sim R^{n-2}$ is
subdominant, thus the Friedmann equation is approximated as
follows,
\begin{equation}
\label{motion1aapprox} \centering \frac{3\lambda n
H^2}{R_0}\left(\frac{R}{R_0}\right)^{n-1}\simeq \frac{\lambda
(n-1)}{2}\left(\frac{R}{R_0}\right)^{n}+ \kappa^2\left(
\frac{1}{2}\dot\varphi^2+V(\varphi)\right)\, .
\end{equation}
Now assuming that the scalar field is slowly rolling, namely,
\begin{equation}\label{slowrollscalarfield}
\frac{1}{2}\dot{\varphi}^2\ll V(\varphi),\,\,\,\ddot{\varphi}\ll
H\dot{\varphi}\, ,
\end{equation}
in conjunction with the approximation of Eq.
(\ref{ricciscalarapprox}) which are justified during the
inflationary era, the Friedmann equation takes the following form,
\begin{equation}\label{friedmannequationfinal}
H^2=\left(\frac{\kappa^2 V(\varphi)}{\beta}\right)^{\frac{1}{n}}\,
,
\end{equation}
where we introduced the parameter $\beta$ which is defined as
$\beta=\tilde{\lambda}\kappa^{2n-2}12^{n-1}\left(6-3n \right)$.
Accordingly, regarding the Raychaudhuri equation (\ref{motion2a}),
the term $\sim \ddot{f}_R$ is subleading, since
$\ddot{f}_R=f_{RRR}\dot{R}^2+f_{RR}\ddot{R}$ and $f_{RR}\sim
R^{n-2}$, $f_{RRR}\sim R^{n-3}$. The same applies for the term
$H\dot{f}_R\sim R^{n-2}H\dot{R}$, hence the Raychaudhuri equation
takes its final form, which is,
\begin{equation}\label{raychauduri}
-2\dot{H}\simeq \frac{\kappa^2\dot{\varphi}^2}{\gamma H^{2n-2}}\,
,
\end{equation}
where we introduced the parameter $\gamma$ which is defined as
follows $\gamma=\tilde{\lambda}n\kappa^{2n-2}12^{n-1}$. Finally,
from the equation that governs the dynamical evolution of the
scalar field (\ref{motion3a}), in view of the slow-roll conditions
for the scalar field given in Eq. (\ref{slowrollscalarfield}), we
have,
\begin{equation}\label{slowrollscalafield}
\dot{\varphi}\simeq -\frac{V'}{3 H}\, .
\end{equation}
The equations (\ref{friedmannequationfinal}), (\ref{raychauduri})
and (\ref{slowrollscalafield}) constitute the differential
equations that govern the evolution of the scalar field in the
power-law $f(R)$ gravity corrected canonical scalar inflation
theory. We quote these here gathered for convenience,
\begin{equation}\label{friedmannequationfinalfinal}
H^2=\left(\frac{\kappa^2 V(\varphi)}{\beta}\right)^{\frac{1}{n}}\,
,
\end{equation}
\begin{equation}\label{raychaudurifinalfinal}
-2\dot{H}\simeq \frac{\kappa^2\dot{\varphi}^2}{\gamma H^{2n-2}}\,
,
\end{equation}
\begin{equation}\label{slowrollscalafieldfinalfinal}
\dot{\varphi}\simeq -\frac{V'}{3 H}\, .
\end{equation}
Let us proceed to the calculation of the slow-roll indices, which
govern the dynamics of inflation, and for a general $f(R,\varphi)$
theory these are defined as follows \cite{Hwang:2005hb,reviews1},
\begin{equation}
\label{epsilon1}\epsilon_1=-\frac{\dot{H}}{H^2},
\end{equation}
\begin{equation}\label{epsilon2}
\epsilon_2=\frac{\ddot{\varphi}}{H\dot{\varphi}}\, ,
\end{equation}
\begin{equation}\label{epsilon3}
\epsilon_3= \frac{\dot{f}_R}{2 H f_R}
\end{equation}
\begin{equation}\label{epsilon4}
\epsilon_4=\frac{\dot{E}}{2H E}\, ,
\end{equation}
where $E$ is defined as follows,
\begin{equation}\label{epsilonparameterE}
E=f_R+\frac{3\dot{f}_R^2}{2\kappa^2\dot{\varphi}^2}\, .
\end{equation}
Let us calculate the above in detail, starting for the simplest,
which is $\epsilon_3$, and it can be written,
\begin{equation}\label{espapprox}
\epsilon_3=\frac{f_{RR}\dot{R}}{2Hf_R}=\frac{(n-1)\dot{R}}{2H R}\,
,
\end{equation}
and in view of Eq. (\ref{ricciscalarapprox}), the final form of
the slow-roll index $\epsilon_3$ during the slow-roll era is quite
simple and it reads,
\begin{equation}\label{epsilon3final}
\epsilon_3\simeq (1-n)\epsilon_1\, .
\end{equation}
Let us proceed to the slow-roll index $\epsilon_2$, and by using
Eq. (\ref{slowrollscalafieldfinalfinal}), this reads,
\begin{equation}\label{slowrollespiklon2}
\epsilon_2=\frac{\frac{d}{dt}\left(-\frac{V'}{3H}\right)}{H\left(-\frac{V'}{3H}\right)}\,
,
\end{equation}
and after some simple algebra, the slow-roll index $\epsilon_2$
acquires the simple form,
\begin{equation}\label{epsilon2final}
\epsilon_2\simeq
-\frac{\beta^{1/n}V''}{3\kappa^{2/n}V^{1/n}}+\epsilon_1\, .
\end{equation}
Regarding the slow-roll index $\epsilon_1$, by combining Eqs.
(\ref{raychaudurifinalfinal}) and
(\ref{friedmannequationfinalfinal}), this takes the following
simplified form,
\begin{equation}\label{epsilon1final}
\epsilon_1=\frac{\beta^{\frac{n+1}{n}}\kappa^{-\frac{2}{n}}}{9\gamma}\frac{\left(V'\right)^2}{V^{\frac{n+1}{n}}}\,
.
\end{equation}
The most tedious to calculate is $\epsilon_4$, which can be
simplified significantly if we further assume that,
\begin{equation}\label{furtherapprox}
3  f_{RR}^2 \dot{R}^2\ll 2 f_R \dot{\varphi}^2\, ,
\end{equation}
which holds true when $n>3/2$, thus eventually $n$ shall be taking
values in the range $3/2<n<2$. The approximation
(\ref{furtherapprox}) is proven easily to hold true, by simply
making use of the slow-roll assumptions for both the scalar and
the Hubble rate, and of course the analytic form of the $f(R)$
gravity function $\sim R^n$. Eventually, after some algebra, the
slow-roll index $\epsilon_4$ reads,
\begin{equation}\label{epsilon4final}
\epsilon_4\simeq \epsilon_1-n\epsilon_1+9\epsilon_1^2-24 n
\epsilon_1^2+21\, n^2\epsilon_1^2-6n^3\epsilon_1^2\, .
\end{equation}
Thus the dynamics of inflation for the power-law $f(R)$ gravity
corrected canonical scalar inflation theory is captured by Eqs.
(\ref{epsilon3final}), (\ref{epsilon2final}),
(\ref{epsilon1final}) and (\ref{epsilon4final}).


Let us now proceed to the calculation of the spectral index of the
primordial scalar curvature perturbations and of the
tensor-to-scalar ratio. As it was shown in Ref.
\cite{Oikonomou:2020krq}, by solely assuming that the slow-roll
indices satisfy $\dot{\epsilon}_i\ll 1$, $i=1,2,3,4$, the spectral
index takes the form \cite{Hwang:2005hb,reviews1},
\begin{equation}
\label{spectralindex} n_s=
1-\frac{4\epsilon_1-2\epsilon_3+2\epsilon_4}{1-\epsilon_1}, .
\end{equation}
The above expression for the spectral index was derived in Ref.
\cite{Hwang:2005hb} directly from the power spectrum of the
primordial scalar curvature perturbations, by using the assumption
$\dot{\epsilon}_i=0$, $i=1,2,3,4$ for the slow-roll indices.
However, this constraint for the slow-roll indices was proven to
be redundant as was evinced in Ref. \cite{Oikonomou:2020krq} where
only the constraint $\dot{\epsilon}_i\ll 1$, $i=1,2,3,4$ is needed
in order to derive the expression (\ref{spectralindex}) for the
spectral index. Notice that the expression for the spectral index
appearing in Eq. (\ref{spectralindex}) holds true even without the
constraint that the slow-roll indices must satisfy the slow-roll
condition $\epsilon_i\ll 1$, for $i=1,2,3,4$. Let us now turn our
focus on the ratio of the tensor power spectrum $P_T$ over the
scalar power spectrum $P_s$ for a general $f(R,\varphi)$ theory is
defined as follows \cite{reviews1},
\begin{equation}\label{tensorananalytic}
r=\frac{P_T}{P_S}=8 \kappa^2 \frac{Q_s}{f_R}\, ,
\end{equation}
where,
\begin{equation}
\label{qsfrpreliminary} Q_s=\frac{\dot{\varphi}E}{f_R\,
H^2(1+\epsilon_3)^2}\, ,
\end{equation}
and $E$ is defined in Eq. (\ref{epsilonparameterE}). By making use
of Eq. (\ref{raychaudurifinalfinal}), the tensor to scalar ratio
for the action (\ref{action1dse}) acquires the quite simplified
form,
\begin{equation}\label{tensortoscalarratio}
r\simeq \frac{16\epsilon_1}{\left( 1+\epsilon_3
\right)^2}+\frac{48\epsilon_1^2}{\left( 1+\epsilon_3
\right)^2}-\frac{96n\epsilon_1^2}{\left( 1+\epsilon_3
\right)^2}+\frac{48n^2\epsilon_1^2}{\left( 1+\epsilon_3
\right)^2}\, ,
\end{equation}
which can be further simplified if the slow-roll indices satisfy
the slow-roll assumptions during inflation, so when $\epsilon_i\ll
1$, with $i=1,2,3,4$, then,
\begin{equation}\label{tensortoscalarratiofinal1}
r\simeq
16\epsilon_1+48\epsilon_1^2-96n\epsilon_1^2+48n^2\epsilon_1\, .
\end{equation}
The reader might easily recognize that the linear term $\sim
\epsilon_1$ is characteristic to the canonical scalar theory
expression of the tensor-to-scalar ratio, while the quadratic
terms $\sim \epsilon_1^2$ are characteristic to $f(R)$ gravity
theories (actually only the first quadratic term, namely
$48\epsilon_1^2$). Our analysis for a specific power-law scalar
potential, indicates that when the slow-roll indices satisfy
$\epsilon_i\ll 1$, with $i=1,2,3,4$, the two expressions appearing
in Eqs. (\ref{tensortoscalarratio}) and
(\ref{tensortoscalarratiofinal1}) yield indistinguishable results,
and in fact only the linear term in the tensor-to-scalar ratio
dominates. Finally, let us provide an expression for the
$e$-foldings number $N$ as a function of the scalar potential. The
$e$-foldings number is defined as follows,
\begin{equation}\label{efoldingsinitial}
N=\int_{t_i}^{t_f}H
dt=\int_{\varphi_i}^{\varphi_f}\frac{H}{\dot{\varphi}} d \varphi
\, ,
\end{equation}
where $t_i$ and $\varphi$ are the time instance at the first
horizon crossing, which we assume that inflation stars, and the
value of the scalar field at the first horizon crossing
respectively, while $t_f$ and $\varphi_f$ are the time instance
and the value of the scalar field at the end of inflation. Using
Eqs. (\ref{friedmannequationfinalfinal}) and
(\ref{slowrollscalafieldfinalfinal}), the $e$-foldings number can
finally be cast in the following form,
\begin{equation}\label{efoldingsfinal}
N=-\frac{3\kappa^{\frac{2}{n}}}{\beta^{\frac{1}{n}}}\int_{\varphi_i}^{\varphi_f}\frac{V^{\frac{1}{n}}}{V'}d
\varphi \, .
\end{equation}
Thus the phenomenology of an arbitrary scalar potential in the
present framework can be directly examined by using Eqs.
(\ref{epsilon3final}), (\ref{epsilon2final}),
(\ref{epsilon1final}), (\ref{epsilon4final}),
(\ref{spectralindex}), (\ref{tensortoscalarratiofinal1}) and
(\ref{efoldingsfinal}). Thus by finding $\varphi_f$ upon solving
the equation $\epsilon_1(\varphi_f)=1$, and upon using Eq.
(\ref{efoldingsfinal}), one may extract an analytic expression for
$\varphi_i$, and thus by calculating the slow-roll indices
(\ref{epsilon3final}), (\ref{epsilon2final}),
(\ref{epsilon1final}), (\ref{epsilon4final}) for
$\varphi=\varphi_i$, one may calculate the spectral index
(\ref{spectralindex}) and the tensor-to-scalar ratio
(\ref{tensortoscalarratiofinal1}) as functions of the free
parameters of the model and of the $e$-foldings number. In the
next section, we shall analyze in detail the simplest choice one
can make for the scalar field potential, namely that of a
power-law model, and we shall examine in detail the inflationary
phenomenology of the model. In principle one may use arbitrary
forms for the scalar potential, as long as for the choice of the
potential, the equations $\epsilon_1(\varphi_f)=1$ and Eq.
(\ref{efoldingsfinal}) can analytically be solved in terms of
$\varphi_f$ and $\varphi_i$ respectively.

\section{$R^n$ Corrected Power-law Scalar Potential}

\begin{figure}
\centering
\includegraphics[width=22pc]{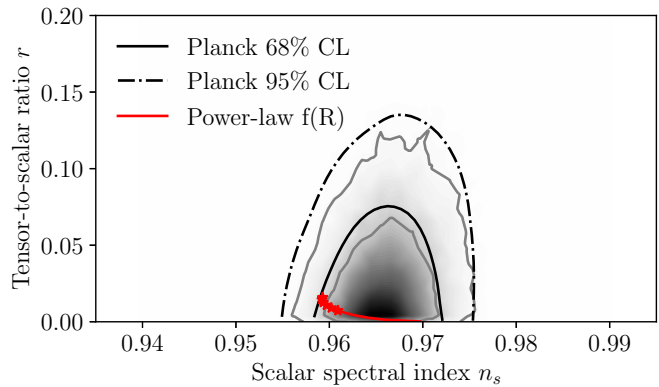}
\caption{Confrontation of the power-law $f(R)$ gravity model (red
curve) with the 2018 Planck constraints for for
$\tilde{\lambda}\sim 10^9$, $V_0\sim 0.5$, $N=60$ and $\mu=0.26$
and with $n$ varying in the range $n=[1.51,1.99]$.}\label{plotnew}
\end{figure}

\begin{figure}
\centering
\includegraphics[width=18pc]{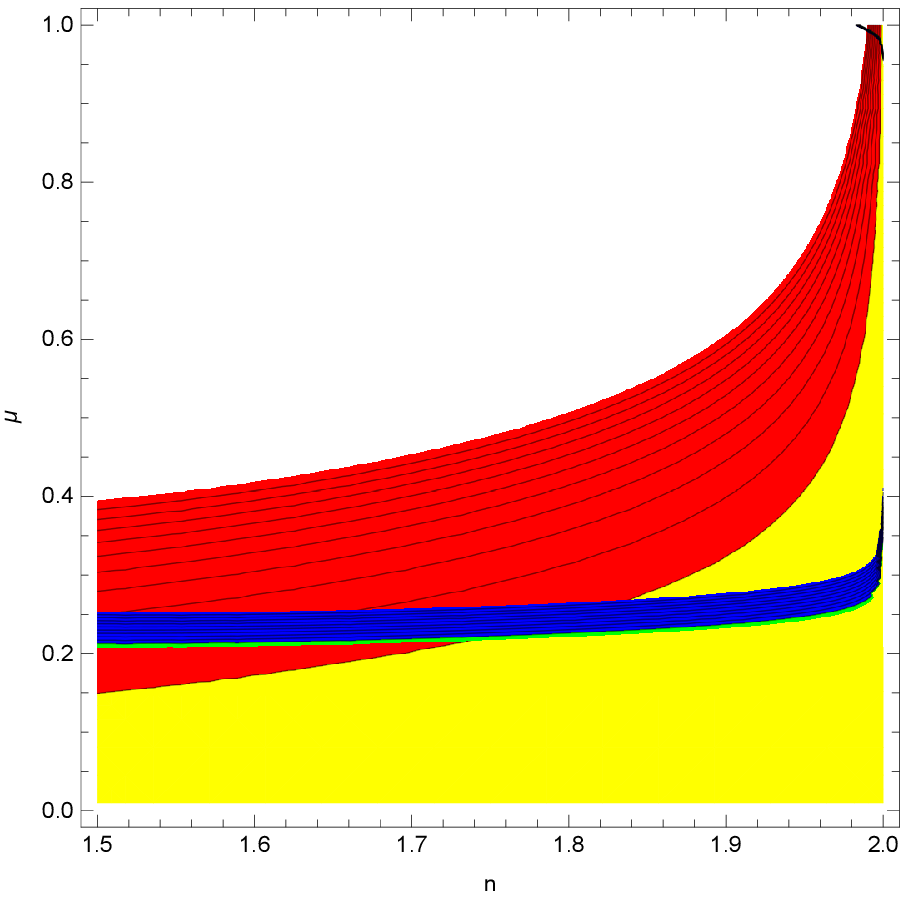}
\includegraphics[width=4pc]{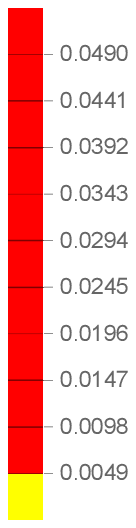}
\includegraphics[width=4pc]{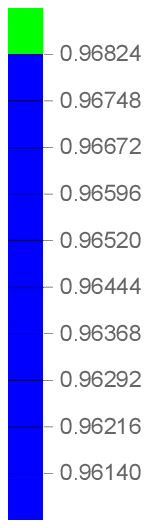}
\caption{Contour plots of the spectral index $n_s$ (blue contours)
and of the tensor-to-scalar ratio $r$ (red and yellow contours)
for $\tilde{\lambda}\sim 10^9$, $V_0\sim 0.5$ and $N=60$. The blue
contours correspond to values of the spectral index in the allowed
by the latest Planck data range $n_s=[0.9607,0.9691]$, while the
red and yellow contours correspond to values of the
tensor-to-scalar ratio $r<0.054$.}\label{plot1}
\end{figure}

In order to exemplify the inflationary phenomenology of the
theoretical framework we developed in the previous section, in
this section we shall analyze in detail the simplest scalar theory
available, by using a power-law scalar potential of the form,
\begin{equation}\label{powerlawpotential}
V(\varphi)=\frac{V_0}{\kappa^4}\left(\kappa \varphi\right)^{\mu}\,
,
\end{equation}
where $V_0$ is a dimensionless constant. Before getting into the
core of the inflationary phenomenology analysis, recall that the
exponent of the $R^n$ term is chosen in the range $3/2<n<2$ (see
below Eq. (\ref{furtherapprox})). Now by substituting the
potential (\ref{powerlawpotential}) in Eqs. (\ref{epsilon3final}),
(\ref{epsilon2final}), (\ref{epsilon1final}),
(\ref{epsilon4final}), (\ref{spectralindex}),
(\ref{tensortoscalarratiofinal1}) and (\ref{efoldingsfinal}) we
may easily study the inflationary phenomenology of the model. Let
us start by quoting the expression for the slow-roll index
$\epsilon_1$, which is,
\begin{equation}\label{epsilon1pot}
\epsilon_1=\frac{\mu ^2 2^{1-\frac{2}{n}} (2-n)^{\frac{1}{n}+1}
\tilde{\lambda}^{1/n} V_0^{1-\frac{1}{n}} \kappa ^{\mu -\frac{\mu
}{n}-2} \varphi ^{\mu -\frac{\mu }{n}-2}}{n }\, ,
\end{equation}
so the equation $\epsilon_1(\varphi_f)=1$ yields the following
value of the scalar field at the end of the inflationary era,
\begin{equation}\label{finalphi}
\varphi_f=\left(\frac{2^{\frac{2}{n}-1} (2-n)^{-\frac{1}{n}-1} n
\tilde{\lambda} ^{-1/n} V_0^{\frac{1}{n}-1} \kappa ^{-\mu
+\frac{\mu }{n}+2}}{\mu ^2}\right)^{\frac{n}{-\mu +\mu  n-2 n}}\,
.
\end{equation}
Accordingly, the slow-roll indices $\epsilon_2$, $\epsilon_3$ and
$\epsilon_4$ can be easily evaluated by using Eqs.
(\ref{epsilon2final}), (\ref{epsilon3final}),
(\ref{epsilon4final}) since the slow-roll indices are basically
functions of $\epsilon_1$. Also, by substituting the potential
(\ref{powerlawpotential}) and $\varphi_f$ from equation
(\ref{finalphi}) in the $e$-foldings equation
(\ref{efoldingsfinal}), we can obtain the value of the scalar
field at the beginning of inflation, during the first horizon
crossing. Finally, the spectral index of the primordial curvature
perturbations and the tensor-to-scalar ratio can be evaluated at
the first horizon crossing by using the obtained initial value of
the scalar field, but we do not quote the final relations for
brevity. The model at hand has four free parameters, namely, the
value of the dimensionless parameter $V_0$, the parameters $n$ and
$\mu$ and also the dimensional parameter $\tilde{\lambda}$, so let
us investigate the parameter space in order to see for which
values the viability with the observational data is obtained.
Recall that the latest Planck data \cite{Akrami:2018odb} constrain
the spectral index and the tensor-to-scalar ratio as follows,
\begin{equation}\label{planck1}
n_{s} = 0.9649 \pm 0.0042, \,\,\, r < 0.056 \, .
\end{equation}
For our analysis, we shall use reduced Planck units, for which
$\kappa=1$, so our investigation indicates that the viability of
the power-law potential model (\ref{powerlawpotential}) occurs for
relatively large values of the parameter $\tilde{\lambda}$, of the
order $\tilde{\lambda}\sim \mathcal{O}(10^9)$, while the rest of
the parameters are of the order of unity. For example if we choose
$\tilde{\lambda}\sim 10^9$, $V_0\sim 0.5$, $n=1.81$, $N=60$ and
$\mu=0.26$, we get,
\begin{equation}\label{resultsphenomenology}
n_s=0.961985,\,\,\, r=0.00531895\, ,
\end{equation}
which are in good agreement with the Planck constraints
(\ref{planck1}). Also there is a wide range of values of
parameters for which the compatibility of the model with the
observational data can be achieved, without fine-tuning. With
regard to the parameter $n$, in Fig. \ref{plotnew} we present the
2018 Planck likelihood curves and we show how the power-law $f(R)$
gravity model (red curve) fits the likelihood curves of Planck,
for $n$ chosen in the range $n=[1.51,1.99]$ and for
$\tilde{\lambda}\sim 10^9$, $V_0\sim 0.5$, $N=60$ and $\mu=0.26$.
As it can be seen in Fig. \ref{plotnew} the power-law $f(R)$
gravity model is well fitted in the Planck observational
constraints.
Moreover, by fixing $\tilde{\lambda}\sim 10^9$, $V_0\sim 0.5$,
$N=60$, and allowing $n$ and $\mu$ to vary freely, recalling that
$3/2<n<2$, in Fig. \ref{plot1} we present the contour plots of the
spectral index (blue contours) and the contour of the
tensor-to-scalar ratio (red and yellow contours). For the spectral
index the blue contours correspond to the values in the range
$n_s=[0.9607,0.9691]$, while for the tensor-to-scalar ratio, the
contours correspond to values less than $r\sim 0.054$. As it can
be seen, the simultaneous compatibility of the observational
indices with the Planck data is achieved for a wide range of the
parameters $\mu$ and $n$, without extreme fine-tuning of the
model. The same applies if $V_0$ varies and the only constraint of
the model is that in all cases, the parameter $\tilde{\lambda}$
must take large values of the order $\tilde{\lambda}\sim 10^9$.
The present theoretical framework can easily accommodate an $f(R)$
gravity term that can drive the dark energy era. Indeed, this term
can be a power-law type term $R^{m}$ which can be chosen to be
subleading compared to the term $R^n$ appearing in the action
(\ref{action1dse}), for example by choosing the exponent $m$ to be
less than unity. This term primordially would be subleading then,
thus it would not affect significantly the evolution, however it
would affect in a dominant way the late-time evolution. This type
of models were considered for example in
\cite{Capozziello:2006uv,Odintsov:2019evb,Oikonomou:2020qah}, and
it is always timely to consider power-law higher curvature
corrections in order to account for dark energy effects. In the
same spirit, one could in principle investigate whether the
Swampland criteria are affected by the combined presence of the
power-law $f(R)$ gravity \cite{Benetti:2019smr} and the scalar
field. In our case, the task would be more involved compared to
the pure $f(R)$ gravity case \cite{Benetti:2019smr}, since in our
case the Einstein frame version of the theory would contain two
scalar fields, but this Swampland study would be interesting to
investigate in a future work.

At this point let us discuss in some detail the main motivation
for choosing an $R^n$ gravity in the presence of a single scalar
field, instead of choosing for example the $R^2$ model to be the
dominant curvature term during inflation. Actually, in Ref.
\cite{Gundhi:2020zvb} the scalaron Higgs model was studied. The
starting point was a similar to ours Jordan frame action,
containing the $R^2$ model plus the Higgs scalar field
gravitational action. The approach adopted in Ref.
\cite{Gundhi:2020zvb} was quite inceptive since the Jordan frame
theory was transformed to the Einstein frame thus a two scalar
field Einstein frame theory was obtained. Accordingly the study
was performed in the presence of two scalar field. In our approach
we remained in the Jordan frame for three main reasons which we
quote:

\begin{itemize}

\item Firstly the Jordan frame power law $f(R)$ gravity of the
form $\sim R^n$ with $1<n<2$, in the absence of the scalar field
is not a viable theory, since it fails to generate a
tensor-to-scalar ratio which is compatible with the latest Planck
data. Thus, in contrast to the $R^2$ model, the power-law $R^n$
model is not a viable theory by itself. Below we shall quote in
brief why the vacuum $R^n$ theory without the scalar field is not
a viable theory.

\item Secondly, if we add a scalar field in the Jordan frame, the
Jordan frame power-law $R^n$ theory is revived and as we showed
the combined theory is compatible with the Planck data.

\item Thirdly, the study was performed in such a way so that all
the equations of motion, the observational indices, the Hubble
rate and the slow-roll indices can be expressed in terms of the
scalar field. This is because the dominant contribution of the
power-law $R^n$ gravity basically alters the first Friedmann
equation, which in our case is,
\begin{equation}\label{revision1}
H^2=\left(\frac{\kappa^2 V(\varphi)}{\beta}\right)^{\frac{1}{n}}\,
,
\end{equation}
instead of the usual Friedman equation for the canonical scalar
field theory,
\begin{equation}\label{revision12}
H^2=\frac{\kappa^2 V(\varphi)}{3}\, .
\end{equation}
Thus technically, the contribution of the power-law $f(R)$ is the
altering of the cosmological equations of motion, and the original
Jordan frame $R^n$ scalar field theory can be treated as a single
scalar field theory with different equations of motion compared to
the single scalar field theory. With this technique, the
complicated initial Jordan frame $f(R,\phi)$ theory is technically
transformed to a single scalar field theory with different
equations of motion compared to the ones in the absence of the
power-law $f(R)$ gravity term. This is the difference of our
approach with Ref. \cite{Gundhi:2020zvb}, where a two scalar field
theory was obtained. Let us also note that in our framework, one
may also choose $n=2$, so the $R^2$ model plus scalar field can be
obtained. However, the $R^2$ model is already successful
phenomenologically, in contrast to the power-law model with
$1<n<2$, thus the latter is more motivated to be studied, since in
our approach it becomes revived.
\end{itemize}
Let us now demonstrate in brief the non-viability of the power-law
$f(R)$ gravity model in the absence of the scalar field. Consider,
\begin{equation}\label{polynomialfr}
f(R)=R+\beta R^n\, ,
\end{equation}
for $n\neq 2$. The first Friedman equation of the vacuum $f(R)$
gravity is,
\begin{equation}\label{friedmannewappendix}
3 H^2F=\frac{RF-f}{2}-3H\dot{F}\, ,
\end{equation}
where $F=\frac{\partial f}{\partial R}$. During inflation, by
$F\sim n\beta R^{n-1}$ so the Friedman equation
(\ref{friedmannewappendix}) becomes,
\begin{align}\label{eqnsofmkotionfrpolyappendix}
& 3 H^2n\beta R^{n-1}=\frac{\beta (n-1)R^{n-1}}{2}-3n(n-1)\beta
HR^{n-2}\dot{R}\, ,
\end{align}
and due to the fact that $R=12H^2+6\dot{H}$, which during
inflation becomes at leading order $R\sim 12 H^2$ and $\dot{R}\sim
24H\dot{H}$, the Friedman equation
(\ref{eqnsofmkotionfrpolyappendix}) becomes after some algebra,
\begin{equation}\label{leadingordereqnappendix}
3H^2n\beta \simeq 6\beta (n-1)H^2-6n\beta(n-1)\dot{H}+3\beta
(n-1)\dot{H}\, ,
\end{equation}
which when solved yields,
\begin{equation}\label{hubblefrpolyappendix}
H(t)=\frac{-2n^2+3n-1}{(n-2)t}\, .
\end{equation}
The above Hubble rate describes an accelerating expansion only for
$1.36<n<2$. Using the Hubble rate of Eq.
(\ref{hubblefrpolyappendix}), the slow-roll indices $\epsilon_i$,
$i=1,..,4$ can be calculated for the $f(R)$ gravity model
(\ref{polynomialfr}), and these are,
\begin{equation}\label{epsiloniforfrpoly}
\epsilon_1=\frac{n-2}{1-3n+2n^2},\,\,\,\epsilon_2\simeq
0,\,\,\,\epsilon_3=(n-1)\epsilon_1,\,\,\,\epsilon_4=\frac{n-2}{n-1}\,
,
\end{equation}
Recall the spectral index and the tensor-to-scalar ratio of a
vacuum $f(R)$ gravity are \cite{reviews1,reviews2},
\begin{equation}\label{vacuumspectral}
n_s=1-6\epsilon_1-2\epsilon_4,\,\,\,r=48 \epsilon_1^2\, .
\end{equation}
The only value of the exponent $n$ that can render the spectral
index compatible with the Planck data is $n=1.81$, however for
this value, the tensor-to-scalar ratio becomes $r=0.13$ which is
incompatible with the observational data of Planck
\cite{Akrami:2018odb}. Thus the vacuum power-law Jordan frame
$f(R)$ gravity model without the scalar field is incompatible with
the Planck data. Let us also note that the scalar power-law model
(\ref{powerlawpotential}) is not compatible with the Planck data
for $\mu\leq 2$, thus with our framework we produced a viable
phenomenology combining two models which were incompatible with
observations otherwise.

\section{Conclusions}

In this work we introduced a modified canonical scalar field
inflationary theoretical framework, for which the dominant
curvature term is not the Ricci scalar, as in Einstein-Hilbert
gravity, but a power-law term $\sim R^n$, with $3/2<n<2$. This
dominance of the power-law term over the Einstein-Hilbert term
eventually modifies the field equations and affects significantly
the slow-roll evolution of the canonical scalar field. We examined
how the field equations become when the dominant curvature term is
the power-law one, and by applying the slow-roll conditions for
both the Hubble rate and for the scalar field, we were able to
express the Friedmann and the Raychaudhuri equations as functions
of the scalar field. We calculated the slow-roll indices for the
resulting theory and we demonstrated that when the slow-roll
conditions hold true, these can be expressed in terms of the
slow-roll index $\epsilon_1$. Accordingly, we gave exact
expressions for the spectral index and for the tensor-to-scalar
ratio, which have quite elegant final functional forms, again
expressed in terms of the slow-roll index $\epsilon_1$ and the
second derivative of the scalar potential with respect to the
scalar field. We used a simple power-law scalar potential in order
to examine the phenomenological viability of the model, and we
demonstrated that the model can be compatible with the latest
Planck data. In conclusion, the unified framework of the power-law
$f(R)$ gravity and of the canonical scalar field, makes possible
to obtain phenomenologically viable models which were not viable
in the absence of either the scalar field or the power-law $f(R)$
gravity term. For example the simple power-law $f(R)$ gravity
model is not compatible with the observational data, and also the
power-law scalar model in the absence of the $f(R)$ gravity term,
is not in general compatible with the Planck data, and it is
compatible for very specific values of the scalar field exponent.
Thus the unified theoretical framework we introduced, offers many
possibilities for realizing viable inflationary phenomenologies. A
mentionable feature of our model is that it falls in the same
category as the models studied in Ref. \cite{delCampo:2012qb},
where the Friedmann equation has the general form
$F(H)=\frac{\kappa^2V(\phi)}{3}$ in the slow-roll approximation.
Such generalized form of the Friedmann equation can be obtained
not only by using extra-dimensional frameworks, but also from
leading order curvature terms during the inflationary era, as we
showed with this work.


\begin{thebibliography}{99}



\bibitem{inflation1}
 A.~D.~Linde,
 Lect.\ Notes Phys.\ {\bf 738} (2008) 1
 [arXiv:0705.0164 [hep-th]].

\bibitem{inflation2} D.~S.~Gorbunov and V.~A.~Rubakov,
``Introduction to the theory of the early universe: Cosmological
perturbations and inflationary theory,'' Hackensack, USA: World
Scientific (2011) 489 p;
%


\bibitem{inflation3}A.~Linde,
arXiv:1402.0526 [hep-th];


\bibitem{inflation4}D.~H.~Lyth and A.~Riotto,
Phys.\ Rept.\  {\bf 314} (1999) 1 [hep-ph/9807278].




\bibitem{Akrami:2018odb}
  Y.~Akrami {\it et al.} [Planck Collaboration],
  arXiv:1807.06211 [astro-ph.CO].






\bibitem{Kamionkowski:2015yta}
  M.~Kamionkowski and E.~D.~Kovetz,
  Ann.\ Rev.\ Astron.\ Astrophys.\  {\bf 54} (2016) 227
  doi:10.1146/annurev-astro-081915-023433
  [arXiv:1510.06042 [astro-ph.CO]].






\bibitem{Brandenberger:2012zb}
  R.~H.~Brandenberger,
  arXiv:1206.4196 [astro-ph.CO].



\bibitem{Brandenberger:2016vhg}
  R.~Brandenberger and P.~Peter,
  arXiv:1603.05834 [hep-th].


\bibitem{Battefeld:2014uga}
 D.~Battefeld and P.~Peter,
 Phys.\ Rept.\ {\bf 571} (2015) 1
 [arXiv:1406.2790 [astro-ph.CO]].


\bibitem{Novello:2008ra}
 M.~Novello and S.~E.~P.~Bergliaffa,
 Phys.\ Rept.\ {\bf 463} (2008) 127
 [arXiv:0802.1634 [astro-ph]].


\bibitem{Cai:2014bea}
  Y.~F.~Cai,
  Sci.\ China Phys.\ Mech.\ Astron.\  {\bf 57} (2014) 1414
  doi:10.1007/s11433-014-5512-3
  [arXiv:1405.1369 [hep-th]].



\bibitem{deHaro:2015wda}
 J.~de Haro and Y.~F.~Cai,
 Gen.\ Rel.\ Grav.\ {\bf 47} (2015) no.8, 95
 [arXiv:1502.03230 [gr-qc]].





\bibitem{Lehners:2011kr}
 J.~L.~Lehners,
 Class.\ Quant.\ Grav.\ {\bf 28} (2011) 204004
 [arXiv:1106.0172 [hep-th]].



\bibitem{Lehners:2008vx}
 J.~L.~Lehners,
 Phys.\ Rept.\ {\bf 465} (2008) 223
 [arXiv:0806.1245 [astro-ph]].

\bibitem{Cheung:2016wik}
 Y.~K.~E.~Cheung, C.~Li and J.~D.~Vergados,
 arXiv:1611.04027 [astro-ph.CO].


\bibitem{Cai:2016hea}
  Y.~F.~Cai, A.~Marciano, D.~G.~Wang and E.~Wilson-Ewing,
  Universe {\bf 3} (2016) no.1,  1
  doi:10.3390/universe3010001
  [arXiv:1610.00938 [astro-ph.CO]].


\bibitem{Odintsov:2020vjb}
S.~D.~Odintsov, V.~K.~Oikonomou, F.~P.~Fronimos and
K.~V.~Fasoulakos,
Phys. Rev. D \textbf{102} (2020) no.10, 104042
doi:10.1103/PhysRevD.102.104042 [arXiv:2010.13580 [gr-qc]].




\bibitem{Aad:2012tfa}
G.~Aad \textit{et al.} [ATLAS],
Phys. Lett. B \textbf{716} (2012), 1-29
doi:10.1016/j.physletb.2012.08.020 [arXiv:1207.7214 [hep-ex]].






\bibitem{Hwang:2005hb}
  J.~c.~Hwang and H.~Noh,
  Phys.\ Rev.\ D {\bf 71} (2005) 063536
  doi:10.1103/PhysRevD.71.063536
  [gr-qc/0412126].




\bibitem{Nojiri:2006je}
  S.~Nojiri, S.~D.~Odintsov and M.~Sami,
  Phys.\ Rev.\ D {\bf 74} (2006) 046004
  doi:10.1103/PhysRevD.74.046004
  [hep-th/0605039].




\bibitem{Cognola:2006sp}
  G.~Cognola, E.~Elizalde, S.~Nojiri, S.~Odintsov and S.~Zerbini,
  Phys.\ Rev.\ D {\bf 75} (2007) 086002
  doi:10.1103/PhysRevD.75.086002
  [hep-th/0611198].



\bibitem{Nojiri:2005vv}
  S.~Nojiri, S.~D.~Odintsov and M.~Sasaki,
  Phys.\ Rev.\ D {\bf 71} (2005) 123509
  doi:10.1103/PhysRevD.71.123509
  [hep-th/0504052].


\bibitem{Nojiri:2005jg}
  S.~Nojiri and S.~D.~Odintsov,
  Phys.\ Lett.\ B {\bf 631} (2005) 1
  doi:10.1016/j.physletb.2005.10.010
  [hep-th/0508049].







\bibitem{Satoh:2007gn}
  M.~Satoh, S.~Kanno and J.~Soda,
  Phys.\ Rev.\ D {\bf 77} (2008) 023526
  doi:10.1103/PhysRevD.77.023526
  [arXiv:0706.3585 [astro-ph]].



\bibitem{Bamba:2014zoa}
  K.~Bamba, A.~N.~Makarenko, A.~N.~Myagky and S.~D.~Odintsov,
  JCAP {\bf 1504} (2015) 001
  doi:10.1088/1475-7516/2015/04/001
  [arXiv:1411.3852 [hep-th]].


\bibitem{Yi:2018gse}
  Z.~Yi, Y.~Gong and M.~Sabir,
  Phys.\ Rev.\ D {\bf 98} (2018) no.8,  083521
  doi:10.1103/PhysRevD.98.083521
  [arXiv:1804.09116 [gr-qc]].


\bibitem{Guo:2009uk}
  Z.~K.~Guo and D.~J.~Schwarz,
  Phys.\ Rev.\ D {\bf 80} (2009) 063523
  doi:10.1103/PhysRevD.80.063523
  [arXiv:0907.0427 [hep-th]].


\bibitem{Guo:2010jr}
  Z.~K.~Guo and D.~J.~Schwarz,
  Phys.\ Rev.\ D {\bf 81} (2010) 123520
  doi:10.1103/PhysRevD.81.123520
  [arXiv:1001.1897 [hep-th]].


\bibitem{Jiang:2013gza}
  P.~X.~Jiang, J.~W.~Hu and Z.~K.~Guo,
  Phys.\ Rev.\ D {\bf 88} (2013) 123508
  doi:10.1103/PhysRevD.88.123508
  [arXiv:1310.5579 [hep-th]].



\bibitem{Kanti:2015pda}
  P.~Kanti, R.~Gannouji and N.~Dadhich,
  Phys.\ Rev.\ D {\bf 92} (2015) no.4,  041302
  doi:10.1103/PhysRevD.92.041302
  [arXiv:1503.01579 [hep-th]].


\bibitem{vandeBruck:2017voa}
  C.~van de Bruck, K.~Dimopoulos, C.~Longden and C.~Owen,
  arXiv:1707.06839 [astro-ph.CO].



\bibitem{Kanti:1998jd}
  P.~Kanti, J.~Rizos and K.~Tamvakis,
  Phys.\ Rev.\ D {\bf 59} (1999) 083512
  doi:10.1103/PhysRevD.59.083512
  [gr-qc/9806085].




\bibitem{Pozdeeva:2020apf}
  E.~O.~Pozdeeva, M.~R.~Gangopadhyay, M.~Sami, A.~V.~Toporensky and S.~Y.~Vernov,
  arXiv:2006.08027 [gr-qc].

\bibitem{Fomin:2020hfh}
  I.~Fomin,
  arXiv:2004.08065 [gr-qc].

\bibitem{DeLaurentis:2015fea}
  M.~De Laurentis, M.~Paolella and S.~Capozziello,
  Phys.\ Rev.\ D {\bf 91} (2015) no.8,  083531
  doi:10.1103/PhysRevD.91.083531
  [arXiv:1503.04659 [gr-qc]].


\bibitem{Chervon:2019sey}
  S.~Chervon, I.~Fomin, V.~Yurov and A.~Yurov,
  doi:10.1142/11405



\bibitem{Nozari:2017rta}
  K.~Nozari and N.~Rashidi,
  Phys.\ Rev.\ D {\bf 95} (2017) no.12,  123518
  doi:10.1103/PhysRevD.95.123518
  [arXiv:1705.02617 [astro-ph.CO]].




\bibitem{Odintsov:2018zhw}
  S.~D.~Odintsov and V.~K.~Oikonomou,
  Phys.\ Rev.\ D {\bf 98} (2018) no.4,  044039
  doi:10.1103/PhysRevD.98.044039
  [arXiv:1808.05045 [gr-qc]].


  \bibitem{Kawai:1998ab}
  S.~Kawai, M.~a.~Sakagami and J.~Soda,
  Phys.\ Lett.\ B {\bf 437}, 284 (1998)
  doi:10.1016/S0370-2693(98)00925-3
  [gr-qc/9802033].


\bibitem{Yi:2018dhl}
  Z.~Yi and Y.~Gong,
  Universe {\bf 5} (2019) no.9,  200
  doi:10.3390/universe5090200
  [arXiv:1811.01625 [gr-qc]].


\bibitem{vandeBruck:2016xvt}
  C.~van de Bruck, K.~Dimopoulos and C.~Longden,
  Phys.\ Rev.\ D {\bf 94} (2016) no.2,  023506
  doi:10.1103/PhysRevD.94.023506
  [arXiv:1605.06350 [astro-ph.CO]].


\bibitem{Kleihaus:2019rbg}
  B.~Kleihaus, J.~Kunz and P.~Kanti,
  arXiv:1910.02121 [gr-qc].





\bibitem{Bakopoulos:2019tvc}
  A.~Bakopoulos, P.~Kanti and N.~Pappas,
  Phys.\ Rev.\ D {\bf 101} (2020) no.4,  044026
  doi:10.1103/PhysRevD.101.044026
  [arXiv:1910.14637 [hep-th]].


\bibitem{Maeda:2011zn}
  K.~i.~Maeda, N.~Ohta and R.~Wakebe,
  Eur.\ Phys.\ J.\ C {\bf 72} (2012) 1949
  doi:10.1140/epjc/s10052-012-1949-6
  [arXiv:1111.3251 [hep-th]].






\bibitem{Bakopoulos:2020dfg}
  A.~Bakopoulos, P.~Kanti and N.~Pappas,
  arXiv:2003.02473 [hep-th].


\bibitem{Ai:2020peo}
W.~Ai,
[arXiv:2004.02858 [gr-qc]].



\bibitem{Odintsov:2019clh}
  S.~D.~Odintsov and V.~K.~Oikonomou,
  Phys.\ Lett.\ B {\bf 797} (2019) 134874
  doi:10.1016/j.physletb.2019.134874
  [arXiv:1908.07555 [gr-qc]].



\bibitem{Oikonomou:2020oil}
V.~K.~Oikonomou and F.~P.~Fronimos,
[arXiv:2007.11915 [gr-qc]].

\bibitem{Odintsov:2020xji}
S.~D.~Odintsov, V.~K.~Oikonomou and F.~P.~Fronimos,
Annals Phys. \textbf{420} (2020), 168250
doi:10.1016/j.aop.2020.168250 [arXiv:2007.02309 [gr-qc]].



\bibitem{Oikonomou:2020sij}
V.~K.~Oikonomou and F.~P.~Fronimos,
[arXiv:2006.05512 [gr-qc]].



\bibitem{Odintsov:2020zkl}
S.~D.~Odintsov and V.~K.~Oikonomou,
Phys. Lett. B \textbf{805} (2020), 135437
doi:10.1016/j.physletb.2020.135437 [arXiv:2004.00479 [gr-qc]].


\bibitem{Odintsov:2020sqy}
S.~D.~Odintsov, V.~K.~Oikonomou and F.~P.~Fronimos,
[arXiv:2003.13724 [gr-qc]].




\bibitem{Odintsov:2020mkz}
S.~D.~Odintsov, V.~K.~Oikonomou, F.~P.~Fronimos and
S.~A.~Venikoudis,
Phys. Dark Univ. \textbf{30} (2020), 100718
doi:10.1016/j.dark.2020.100718 [arXiv:2009.06113 [gr-qc]].


\bibitem{Easther:1996yd}
  R.~Easther and K.~i.~Maeda,
  Phys.\ Rev.\ D {\bf 54} (1996) 7252
  doi:10.1103/PhysRevD.54.7252
  [hep-th/9605173].

\bibitem{Antoniadis:1993jc}
  I.~Antoniadis, J.~Rizos and K.~Tamvakis,
  Nucl.\ Phys.\ B {\bf 415} (1994) 497
  doi:10.1016/0550-3213(94)90120-1
  [hep-th/9305025].

\bibitem{Antoniadis:1990uu}
I.~Antoniadis, C.~Bachas, J.~R.~Ellis and D.~V.~Nanopoulos,
Phys.\ Lett.\ B \textbf{257} (1991), 278-284
doi:10.1016/0370-2693(91)91893-Z




\bibitem{Kanti:1995vq}
P.~Kanti, N.~Mavromatos, J.~Rizos, K.~Tamvakis and E.~Winstanley,
Phys. Rev. D \textbf{54} (1996), 5049-5058
doi:10.1103/PhysRevD.54.5049 [arXiv:hep-th/9511071 [hep-th]].



\bibitem{Kanti:1997br}
P.~Kanti, N.~Mavromatos, J.~Rizos, K.~Tamvakis and E.~Winstanley,
Phys. Rev. D \textbf{57} (1998), 6255-6264
doi:10.1103/PhysRevD.57.6255 [arXiv:hep-th/9703192 [hep-th]].











\bibitem{reviews1}
 S.~Nojiri, S.~D.~Odintsov and V.~K.~Oikonomou,
  Phys.\ Rept.\  {\bf 692} (2017) 1
  [arXiv:1705.11098 [gr-qc]].

\bibitem{reviews2}


 S. Capozziello, M. De Laurentis,
   Phys.\ Rept.\  {\bf 509}, 167 (2011);\\
V. Faraoni and S. Capozziello, The landscape beyond Einstein
gravity, in Beyond Einstein Gravity 828 (Springer, Dordrecht,
2010), Vol. 170, pp.59-106.



   \bibitem{reviews3}

S. Nojiri, S.D. Odintsov,
   Phys.\ Rept.\  {\bf 505}, 59 (2011);




\bibitem{reviews4}

A.~de la Cruz-Dombriz and D.~Saez-Gomez,
  Entropy {\bf 14} (2012) 1717
  [arXiv:1207.2663 [gr-qc]].


\bibitem{dimopoulos}Introduction to Cosmic Inflation and Dark
Energy, Konstantinos Dimopoulos, CRC Press 2020



\bibitem{Aghanim:2018eyx}
N.~Aghanim \textit{et al.} [Planck],
Astron. Astrophys. \textbf{641} (2020), A6
doi:10.1051/0004-6361/201833910 [arXiv:1807.06209 [astro-ph.CO]].


\bibitem{Caldwell:2003vq}
R.~R.~Caldwell, M.~Kamionkowski and N.~N.~Weinberg,
Phys. Rev. Lett. \textbf{91} (2003), 071301
doi:10.1103/PhysRevLett.91.071301 [arXiv:astro-ph/0302506
[astro-ph]].


\bibitem{Oikonomou:2020krq}
V.~Oikonomou, ``Rectifying an Inconsistency in $F(R)$ Gravity
Inflation,'' European Physics Letters in press, [arXiv:2004.10778
[gr-qc]].




\bibitem{Gundhi:2020zvb}
A.~Gundhi and C.~F.~Steinwachs,
[arXiv:2011.09485 [hep-th]].


\bibitem{Capozziello:2006uv}
S.~Capozziello, V.~F.~Cardone and A.~Troisi,
JCAP \textbf{08} (2006), 001 doi:10.1088/1475-7516/2006/08/001
[arXiv:astro-ph/0602349 [astro-ph]].



\bibitem{Odintsov:2019evb}
S.~D.~Odintsov and V.~K.~Oikonomou,
Phys. Rev. D \textbf{99} (2019) no.10, 104070
doi:10.1103/PhysRevD.99.104070 [arXiv:1905.03496 [gr-qc]].


\bibitem{Oikonomou:2020qah}
V.~K.~Oikonomou,
Phys. Rev. D \textbf{103} (2021) no.4, 044036
doi:10.1103/PhysRevD.103.044036 [arXiv:2012.00586 [astro-ph.CO]].


\bibitem{Benetti:2019smr}
M.~Benetti, S.~Capozziello and L.~L.~Graef,
Phys. Rev. D \textbf{100} (2019) no.8, 084013
doi:10.1103/PhysRevD.100.084013 [arXiv:1905.05654 [gr-qc]].


\bibitem{delCampo:2012qb}
S.~del Campo,
JCAP \textbf{12} (2012), 005 doi:10.1088/1475-7516/2012/12/005
[arXiv:1212.1315 [astro-ph.CO]].




\end{thebibliography}
\end{document}